\documentclass[twoside]{ilcws10}
\usepackage[latin1]{inputenc}
\usepackage[dvips]{graphicx,epsfig,color}
\usepackage{wrapfig,rotating}
\usepackage{amssymb,amsmath,array}
\usepackage{subfigure}

\begin{document}
\title{
Probing Anomalous Top-Gluon Couplings at Colliders} 
\author{Pratishruti Saha
\vspace{.3cm}\\
Department of Physics and Astrophysics, University of Delhi\\
Delhi-110007, India
}

\maketitle

\begin{abstract}
Anomalous chromomagnetic and chromoelectric dipole moments of the top quark
may arise from various high scale theories. We carry out a model independent study
of such interactions focusing on the limits that can be obtained from current Tevatron
data and the improvements that may be possible at the LHC or at a future Linear Collider.
\end{abstract}

\section{Introduction}

The top quark was discovered at the Tevatron proton-antiproton collider
at Fermilab in 1995~\cite{top_discovery}. The study of its properties continues even today.
The top is the heaviest particle in the Standard Model (SM)
with a mass $\sim$175 GeV which differs widely from those of the other
fundamental fermions. This seems to suggest that the top quark may have a role to play in
electroweak symmetry breaking and prompts us to question whether the top quark
has couplings different from and in addition to those of the other quarks.

Various anomalous couplings of the top have been discussed in
Ref.~\cite{eff_terms}. Of these, the ones that pertain to the QCD-sector
form the subject of this study. 
Large anomalous couplings may arise in a plethora of models~\cite{relevant_models},
contributing to higher order corrections to the $ttg$ vertex.
In a model independent framework, the lowest-dimensional anomalous coupling
of the top with the gluon can be parametrized by extra terms in the
interaction Lagrangian of the form 
\begin{equation}
{\cal L}_{int} \ni 
\frac{g_s}{\Lambda} \, F^{\mu\nu}_a  \; 
\bar t  \sigma_{\mu\nu} (\rho + i \, \rho' \, \gamma_5) \, T_a \, t 
\label{lagrangian}
\end{equation}
where $\Lambda$ denotes the scale of the effective theory.  While
$\rho$ represents the anomalous chromomagnetic dipole moment of the
top, $\rho'$ indicates the presence of a ($CP$-violating)
chromoelectric dipole moment. Within the SM, $\rho'$ is non-zero only
at the three-loop level and is, thus, tiny. $\rho$, on the other hand,
receives a contribution at the one-loop level and is ${\cal
  O}(\alpha_s/\pi)$ for $\Lambda \sim m_t$. The evidence for a larger
$\rho$ or $\rho'$ would be a strong indicator of new physics
lurking nearby. Whereas both $\rho$ and $\rho'$ can, in general, be
complex, note that any imaginary part thereof denotes absorptive
contributions and would render the Lagrangian non-Hermitian. We desist
from considering such a possibility.

The phenomenological consequences of such anomalous couplings have been 
considered earlier in Ref.~\cite{previous}. We reopen the issue
in light of the improved measurements of top quark mass and $t\bar t$
cross-section and the first reported measurement of $t\bar t$ invariant mass.

\section{Hadron Collider Prospects}
The inclusion of a chromomagnetic moment term leads to a modification
of the vertex factor for the usual $ttg$ interaction to
$ig_s[\gamma^{\alpha} +
 (2 \, i \, \rho / \Lambda) \, \sigma^{\alpha\mu}k_{\mu}]T^a$ where 
$k$ is the momentum of the gluon coming
into the vertex.  An additional quartic interaction involving two top
quarks and two gluons is also generated
with the corresponding vertex factor being
$(2 \, i \,g_s^2 \, \rho / \Lambda) \, f_{abc}\sigma^{\alpha\beta}T^c$. 
The changes in the presence of the 
chromoelectric dipole moment term are analogous, 
with $\rho$ above being replaced by $ (i \, \rho' \, \gamma_5)$.

At a hadron collider, the leading order 
contributions to $t\bar t$ production come from
the $q\bar q \rightarrow t\bar t$ and $gg \rightarrow t\bar t$
sub-processes. Detailed expressions for the differential cross-sections
can be found in Ref.~\cite{chromo_top} as well as Ref.~\cite{previous}.

Using these results, we compute the
expected $t\bar t$ cross-section at the Tevatron and the
LHC. We use the CTEQ6L1 parton distribution
sets~\cite{CTEQ} with $m_{t}$ as the scale for both factorization as
well as renormalization. For a consistent comparison with the 
cross-section measurement reported by the CDF collaboration~\cite{CDF_top_csec},
we use $m_{t} = 172.5$ GeV for the Tevatron analysis. For the LHC analysis, 
though, we use the updated value of $m_{t} = 173.1$ GeV, obtained from 
combined CDF+D\O{} analysis~\cite{CDF-D0}. To incorporate the
higher order corrections absent in our leading order results, we use
the $K$-factors at the NLO+NLL level\footnote{In the absence of a
  similar calculation incorporating anomalous dipole moments, we use
  the same \mbox{$K$-factor} as obtained for the SM case. 
  While this is not entirely accurate, given the fact that the color structure
  is similar the error associated with this approximation is not expected to be large.}
calculated by Cacciari et. al.~\cite{Cacciari}.  
Once this is done, the theoretical errors in
the calculation owing to the choice of PDFs and scale are approximately
7-8\% for the Tevatron and 9-10\% for the LHC~\cite{Cacciari}.
However, the estimates reported for the LHC operating at 7 TeV, are only
leading order ones since NLO calculations for these energies 
are, so far, unavailable.

\subsection{Tevatron Results} 

\begin{figure}[!htbp]
\centering
\subfigure[]
{
\includegraphics[scale=0.75]{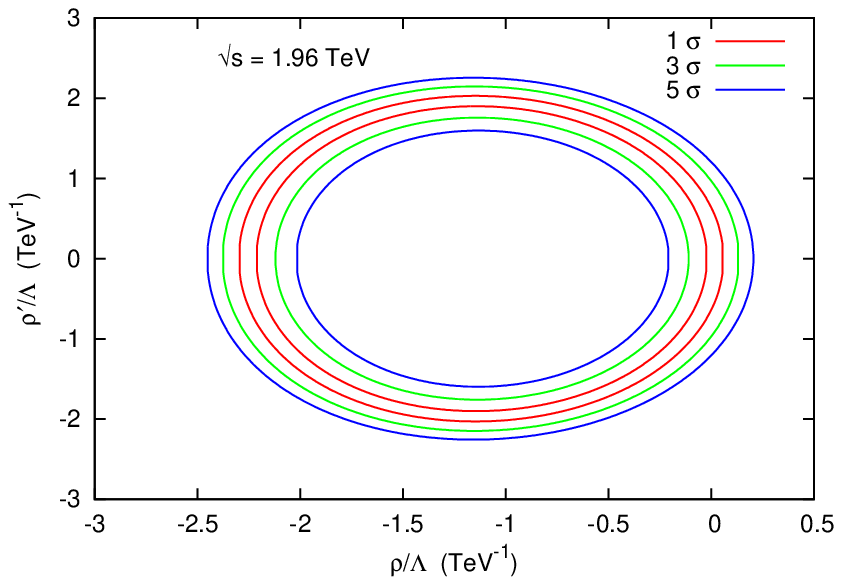}
}
\subfigure[]
{
\includegraphics[scale=0.75]{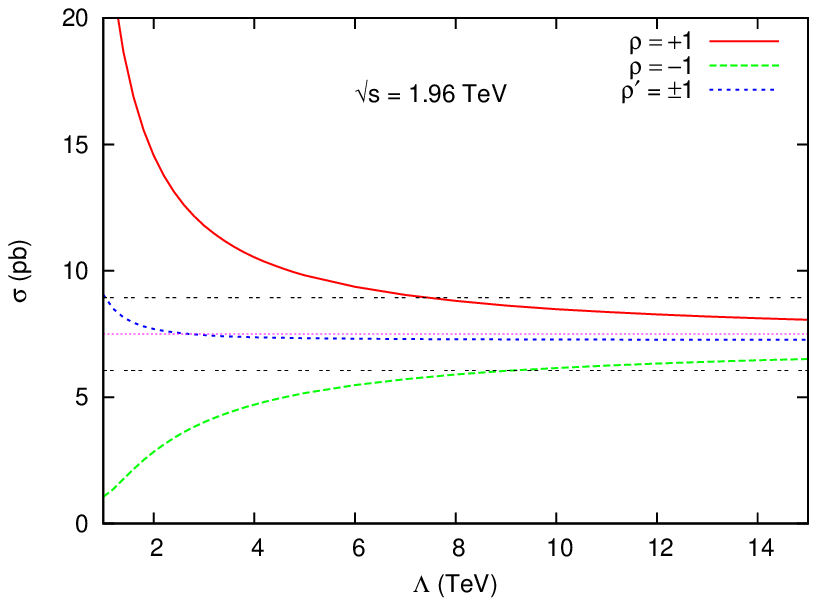}
}
\caption{\em {\em (a)} 
The region in ($\rho/\Lambda$)-($\rho'/\Lambda$) plane allowed
by the Tevatron data~{\em\cite{CDF_top_csec}} at the 1-$\sigma$, 3-$\sigma$
and 5-$\sigma$ level. {\em (b)} 
$t\bar t$ production rates for the Tevatron ($\sqrt{s}$ = {\em 1.96 TeV)}.
The horizontal lines denote the CDF central value and the
3-$\sigma$ interval~{\em\cite{CDF_top_csec}}.}
\label{fig:contour}
\end{figure}

At the Tevatron, the dominant contribution accrues from the $q \bar q$
initial states, even on the inclusion of the dipole moments.  
Fig.\ref{fig:contour}($a$) displays the parameter space that is
still allowed by the Tevatron data, namely~\cite{CDF_top_csec}
\begin{equation}
  \sigma_{t\bar t}(m_t = 172.5~{\rm GeV}) = (7.50 \pm 0.48) \, {\rm pb}\,.
\end{equation}
The central region of the plot shows that the data
allows for large values of dipole moments. This is essentially
due to cancellations 
between various terms contributing to the cross-section. 

Having seen the extent to which cancellations may, in principle, be
responsible for hiding the presence of substantial dipole moments, we
now restrict ourselves to the case where only one of $\rho$ and
$\rho'$ may be non-zero. If only one of the two couplings is to be 
non-zero, we may rescale $\rho, \rho' = 0, \pm 1$ and, thus, 
reduce the parameter space to one
dimension ($\Lambda$). $\rho' = \pm 1$ are equivalent as the the cross-section
only depends on even powers of $\rho'$.

Fig.\ref{fig:contour}($b$) exhibits the corresponding dependence of
the total cross-section at the Tevatron on $\Lambda$ for various
combinations of $(\rho, \rho')$. 
It can be seen that $\Lambda \lesssim 7400~{\rm GeV}$ can be ruled out at 99\%
confidence level for the $\rho = +1$ case. For $\rho = -1$ on the
other hand, $\Lambda \lesssim 9000~{\rm GeV}$ can be ruled out at the same
confidence level. One expects similar sensitivity for $\rho=+1$ and $\rho=-1$.
The difference essentially owes its origin to the
slight discrepancy between the SM expectations (as computed with our
choices) and the experimental central value. The sensitivity to chromoelectric moment
is low. This is understandable as the corresponding contribution
is suppressed by at least $\Lambda^2$.

\begin{figure}[!htbp]
\centering
\subfigure[]
{
\includegraphics[scale=0.75]{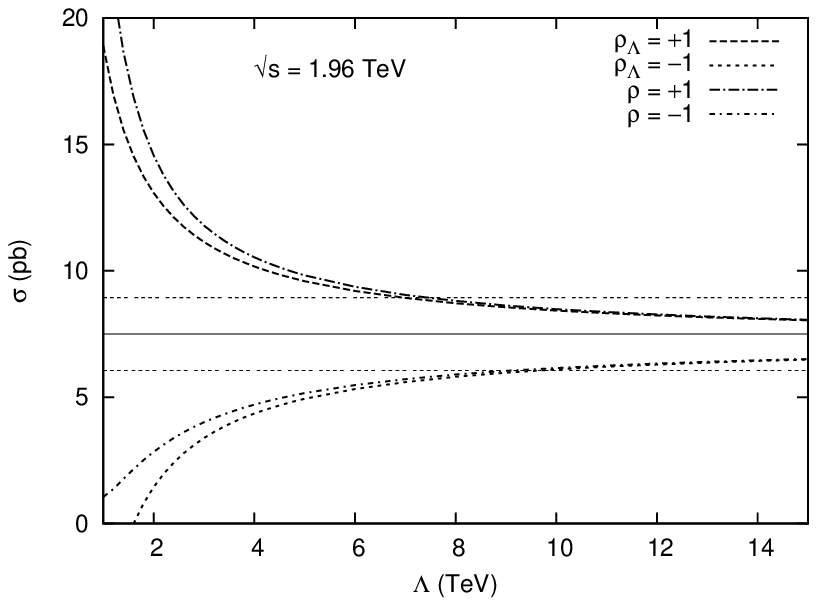}
}
\subfigure[]
{
\includegraphics[scale=0.75]{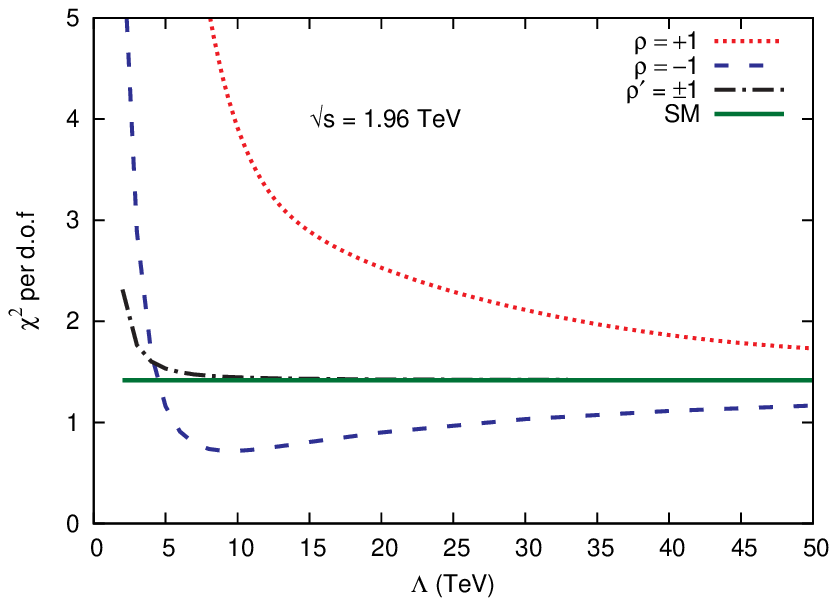}
}
\caption{\em 
{\em (a)} 
Comparison of production rates obtained at the Tevatron with truncated cross-sections 
(up to ${\cal O}(\Lambda^{-1})$; denoted by subscript $\Lambda$ in the key) and 
full cross-sections(all orders in $\Lambda$).
{\em (b)} 
$\chi^2$ per degree of freedom obtained by fitting the $m_{t\bar t}$ spectrum.
}
\label{fig:chisq}
\end{figure}

The cross-sections considered above depend on powers of $\Lambda$ upto
$\Lambda^4$. However, the Lagrangian considered in Eqn.\ref{lagrangian}
contains only the lowest dimensional anomalous operators of an effective
theory. Higher dimensional operators~\cite{dim_6}, if included in the 
Lagrangian, could change the behaviour of the cross-sections and hence the
conclusions drawn from Fig.\ref{fig:contour}($b$). A closer examination of
this issue (see Fig.\ref{fig:chisq}($a$)) reveals that that were we to neglect
${\cal O}(\Lambda^{-2})$ terms, the shape of the curves would
indeed change but the limits on $\Lambda$ for either of $\rho = \pm 1$ would hardly alter.

Yet another measurement reported by the Tevatron is the invariant mass distribution~\cite{CDF_mtt}.
This data can be used to put further constraints on values of $\rho$ and $\rho'$.
In the reported measurement, the first bin which extends in the range 0-350 GeV 
also has a non-zero number of events, an artefact of experimental errors
associated with the reconstruction of the $t\bar t$ events. 
For our analysis, we exclude this bin. Further, we normalize
the our calculated $m_{t\bar t}$ distribution so that for the SM case it matches the CDF simulation.
As a statistic, we consider a $\chi^2$ defined through
\[
   \chi^2 = \sum_{i = 2}^9 \left(\frac{\sigma^{\rm th}_i - \sigma^{\rm obs}_i}{\delta \sigma_i}
                          \right)^2
\]
where the sum runs over the bins and $\sigma^{\rm th}_i$ is the number 
of events expected in a given theory (defined by the values of 
$\rho, \rho', \Lambda$)
in a particular bin. $\sigma^{\rm obs}_i$ and $\delta \sigma_i$  
are the observed event numbers and the errors therein. The
$\chi^2$ values thus obtained are plotted as function of $\Lambda$
in Fig.\ref{fig:chisq}($b$). 

It is interesting to note that the $\rho = -1$ case gives a
better fit than the SM, over a
large range of $\Lambda$ values while $\rho =  +1$ is now strongly 
disfavoured for much higher values of $\Lambda$. 
Even for the chromoelectric moment case 
($\rho' \neq 0$), the increase in sensitivity is evident. 
However, in all of this, we wish to
tread with caution.  This distribution has been constructed on the
basis of only 2.7 $fb^{-1}$ of data.  Robust limits may be obtained
once more statistics has been accumulated and a more realistic
simulation, with the inclusion of the effects of dipole moment
terms, has been carried out.

\subsection{LHC Sensitivity}

\begin{figure}[!htbp]
\centering
\subfigure[]
{
\includegraphics[width=1.7in,height=2in]{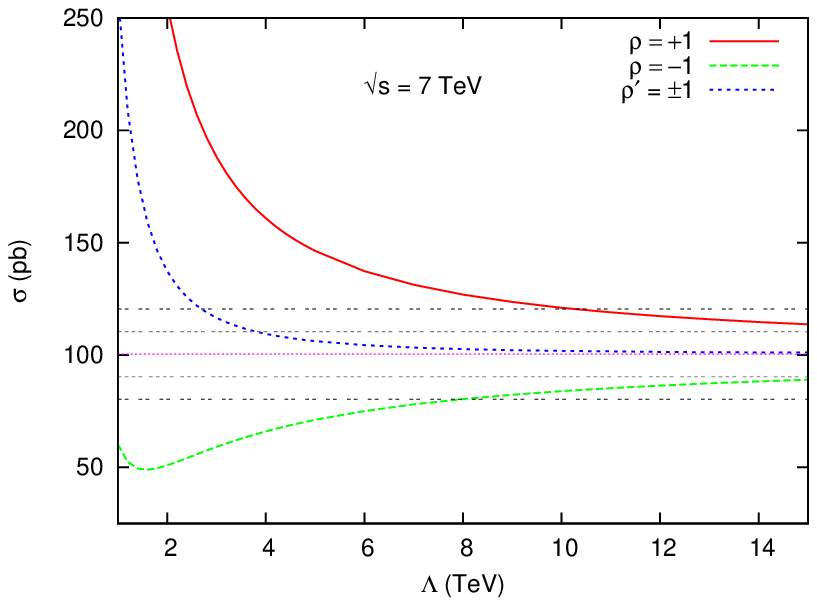}
}
\subfigure[]
{
\includegraphics[width=1.7in,height=2in]{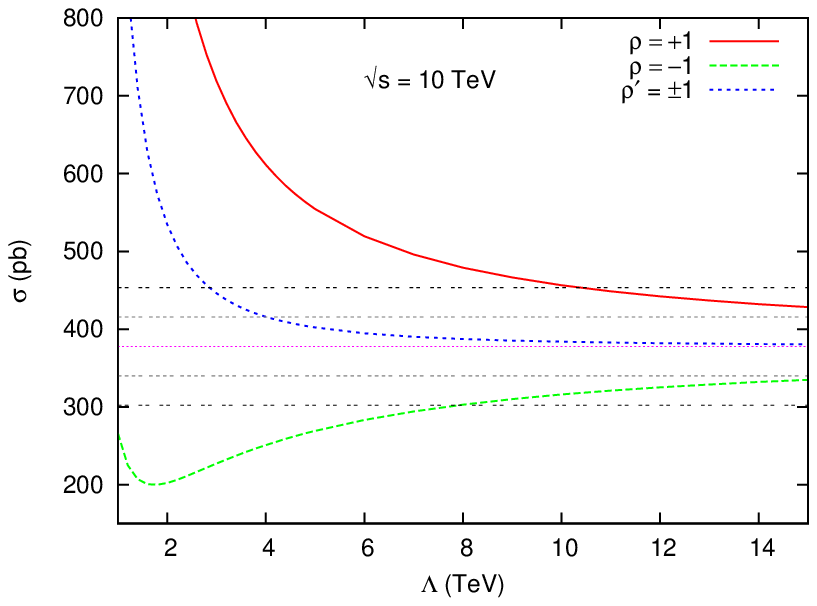}
}
\subfigure[]
{
\includegraphics[width=1.7in,height=2in]{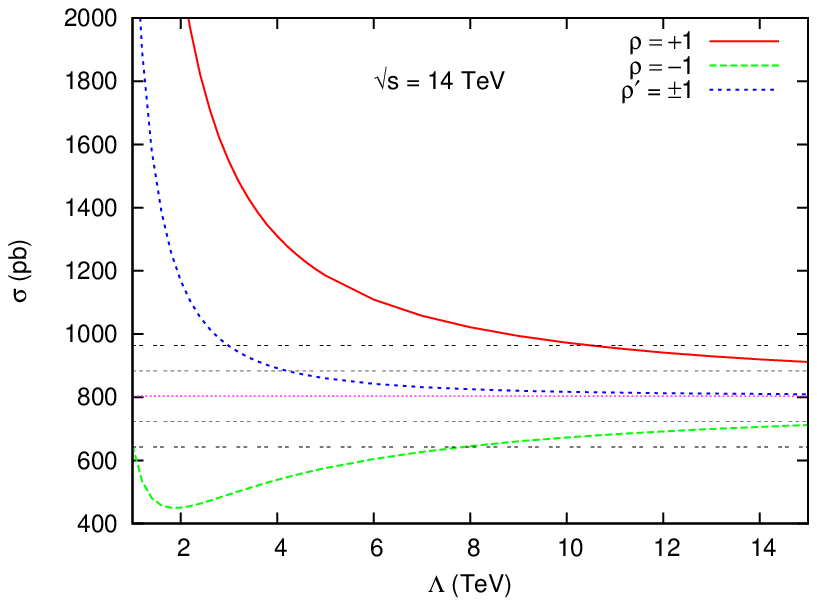}
}
\caption{\em $t\bar t$ production rates for the LHC as a function of
  the new physics scale $\Lambda$. Panels from left to right
  correspond to $\sqrt{s}$ = {\em 7, 10, 14 TeV}. 
  The horizontal lines show the SM expectation and the 10\% and
  20\% intervals as estimates of errors in the measurement~{\em \cite{CMS-PAS}}.}
\label{fig:tot_LHC}
\end{figure}

At the LHC, the $gg$ flux dominates, especially at 
smaller $\hat s$ values. In Fig.\ref{fig:tot_LHC}, we present the 
cross-sections at the LHC for various values of 
the proton-proton center-of-mass energy $\sqrt{s}$. 
In the absence of any data, we can only compare these with the 
SM expectations and the estimated errors~\cite{CMS-PAS}.

For non-zero $\rho'$, an early run
of the LHC with $\sqrt{s}$ = 7 TeV (Fig.\ref{fig:tot_LHC}$a$) 
would be sensitive to $\Lambda \lesssim 2700~{\rm GeV}$.  
The improvement of the sensitivity with the machine energy is marginal
at best.
For $\rho = +1$, naively a sensitivity up to about
$\Lambda \sim 10~{\rm TeV}$ could be expected.
For $\rho = -1$, on the other hand, it appears
that the best that the LHC can do is to rule out (for $\rho = -1$)
$\Lambda \lesssim 8~{\rm TeV}$. This, however, should be compared with the Tevatron
results which have already ruled out $\Lambda \lesssim 9~{\rm TeV}$.
 
\subsection{Summary of Limits from Hadron Colliders}
Rephrasing the above results in terms of notation commonly used in literature: 
\begin{center}
$\dfrac{1}{\Lambda}(\rho + i\rho') \longleftrightarrow \dfrac{1}{2m_t}(\kappa + i\tilde\kappa)$
\hspace{15pt} : \hspace{15pt}
$-0.038 \leq \kappa \leq$ 0.034 \hspace{5pt} and \hspace{5pt} $|\tilde\kappa| \leq$ 0.12
\end{center}

\section{Linear Collider Prospects}

\begin{figure}[!htbp]
\centering
\subfigure[]
{
\includegraphics[width=2.6in,height=2.6in,angle=-90]{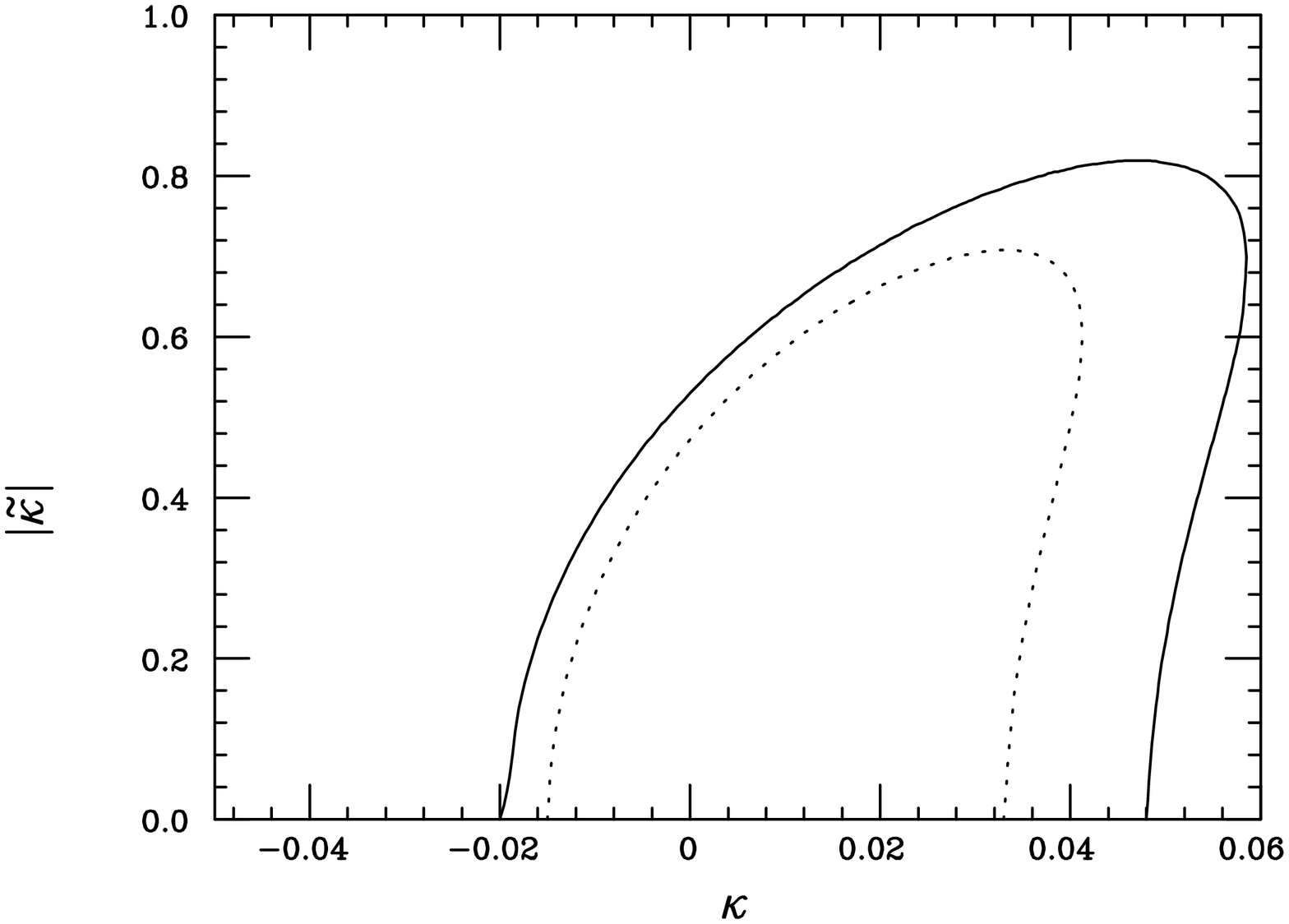}
}
\subfigure[]
{
\includegraphics[width=2.6in,height=2.6in,angle=-90]{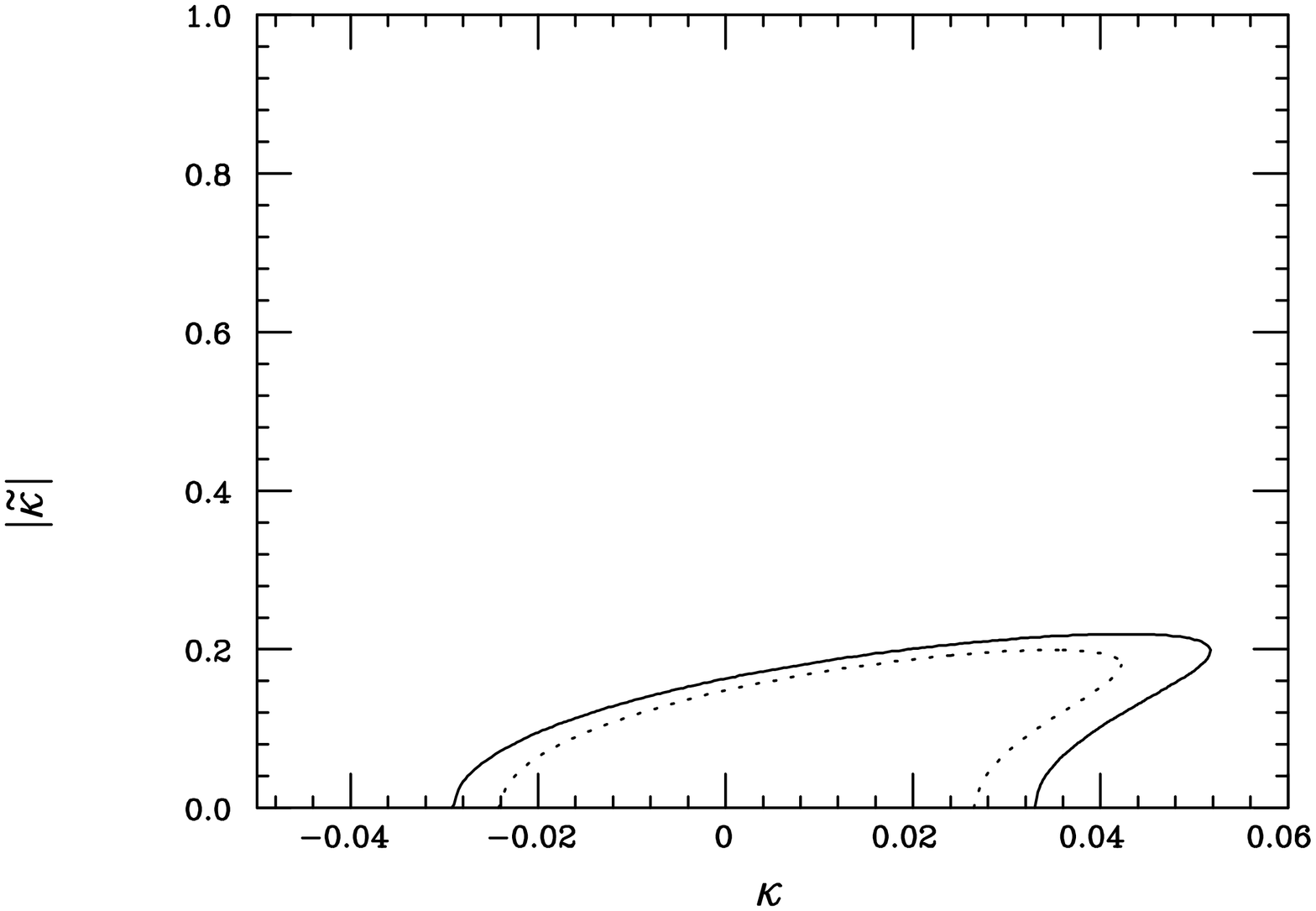}
}
\caption{\em Reproduced from Ref.~{\em \cite{Rizzo_96}}. Shows the 95\% CL allowed region for
{\em (a)} 
$\sqrt{s}$ = {\em 500 GeV} ; ${\cal L}$ = {\em 50 $fb^{-1}$}(solid), {\em 100 $fb^{-1}$}(dotted) ; $E_g^{min}$ = {\em 25 GeV}.
{\em (b)}
$\sqrt{s}$ = {\em 1 TeV} ; ${\cal L}$ = {\em 100 $fb^{-1}$}(solid), {\em 200 $fb^{-1}$}(dotted) ; $E_g^{min}$ = {\em 25 GeV}.
}
\label{fig:Rizzo}
\end{figure}

An electron-positron collider would be the ideal ground for probing anomalous
electroweak couplings of the top quark.
However, anomalous top-gluon couplings would play a role in the process 
$e^+e^- \rightarrow t\bar tg$. This has been studied in Refs.~\cite{Rizzo_94} and \cite{Rizzo_96},
where, it was shown that the energy distribution of the gluon is sensitive to such anomalous couplings. 
Limits on couplings were obtained by fitting the energy spectrum of the gluon
assuming that there is no excess in the total production cross-section. Some of the results
from Ref.~\cite{Rizzo_96} are shown in Fig.\ref{fig:Rizzo}. 

Considering only one of $\kappa$ and $\tilde\kappa$ to be non-zero at a time,
the dotted curve in Fig.\ref{fig:Rizzo}($a$) implies \mbox{-0.015 $\leq \kappa \leq$ 0.033} and 
\mbox{$|\tilde\kappa| \leq$ 0.47}.
With an increase in center-of-mass energy and luminosity, this limit may be improved
as indicated by Fig.\ref{fig:Rizzo}($b$). Here, the dotted curve leads to
\mbox{-0.024 $\leq \kappa \leq$ 0.026} and \mbox{$|\tilde\kappa| \leq$ 0.14}.
Comparing this to the limits expected from hadron colliders listed in the previous section, it can be seen that,
at a linear collider, better sensitivity may be expected for $\kappa$ but not for $\tilde\kappa$.

At a linear collider, there also exists the possibility of collisions using a polarized beam.
This too was studied in Ref.~\cite{Rizzo_96}. However it was found that, using a polarized beam 
does not yield better limits on either chromomagnetic or chromoelectric dipole moments
as compared to what can be obtained with unpolarized beams. 

\section*{\begin{center}Acknowledgement\end{center}}
The author would like to thank the organisers of LCWS10 for support and hospitality
and the \mbox{ILC-India Forum} for funding the trip to Beijing to attend this conference.


\begin{footnotesize}


\end{footnotesize}



\begin{thebibliography}{99}

\bibitem{top_discovery} 
F.~Abe {\it et~al.}, Phys. Rev. Lett. {\bf 74} 2626 (1995);
S.~Abachi {\it et~al.}, Phys. Rev. Lett. {\bf 74} 2632 (1995).

\bibitem{eff_terms} 
J.A.~Aguilar-Saavedra, Nucl. Phys. {\bf B812} 181 (2009).

\bibitem{relevant_models}
C.T.~Hill, Phys. Lett. {\bf B266} 419 (1991); 
Phys. Lett. {\bf B345} 483 (1995);
W.~Buchmuller and D.~Wyler, Nucl. Phys. {\bf B268} 621 (1986);
C.~Arzt {\it et~al.}, Nucl. Phys. {\bf B433} 41 (1995);
N.~Arkani-Hamed {\it et~al}, Phys. Lett. {\bf B513} 232 (2001);
C.~Csaki {\it et~al.}, Phys. Rev. {\bf D67} 115002 (2003);  
Phys. Rev. {\bf D68} 035009 (2003);
J.L.~Hewett {\it et~al.}, JHEP {\bf 0310} 062 (2003);
M.C.~Chen and S.~Dawson, Phys. Rev. {\bf D70} 015003 (2004);
W.~Kilian and J.~Reuter, Phys. Rev. {\bf D70} 015004 (2004);
Z.~Han and W.~Skiba, Phys. Rev. {\bf D71} 075009 (2005);
T.~Appelquist {\it et~al.}, Phys. Rev. {\bf D64} 035002 (2001);
I.~Antoniadis, Phys. Lett. {\bf B246} 377 (1990);
N.~Arkani-Hamed and M.~Schmaltz, Phys. Rev. {\bf D61} 033005 (2000);
R.~Barbieri {\it et~al.}, Phys. Rev. {\bf D63} 105007 (2001);
G.~Cacciapaglia {\it et~al.}, Nucl. Phys. B {\bf B634} 230 (2002);
J.L.~Hewett and T.G.~Rizzo, Phys. Rept. {\bf 183} 193 (1989);
G.~Bhattacharyya {\it et~al.}, Phys. Lett. {\bf B355} 193 (1995).

\bibitem{previous}
D.~Atwood {\it et~al.}, Phys. Rev. Lett. {\bf 69} 2754 (1992);
T. G. Rizzo, DPF Conf. 717 (1994);
D.~Atwood {\it et~al.}, Phys. Rev. {\bf D52} 6264 (1995);
P.~Haberl {\it et~al.}, Phys. Rev. {\bf D53} 4875 (1996); 
K.~Cheung, Phys. Rev. {\bf D53} 3604 (1996);
T.G.~Rizzo, Proceedings of 1996 DPF/DPB Summer Study on New Directions
for High-Energy Physics (Snowmass 96);
S.Y.~Choi {\it et~al.}, Phys. Lett. {\bf B415} 67 (1997);
B.~Grzadkowski {\it et~al.}, Phys. Lett. {\bf B415} 193 (1997);
B.~Lampe, Phys. Lett. {\bf B415} 63 (1997);
H.Y.~Zhou, Phys. Rev. {\bf D58} 114002 (1998);
K.~Hikasa {\it et~al.}, Phys. Rev. D {\bf D58} 114003 (1998);
K.~Ohkuma, arXiv:hep-ph/0105117 (2001);
R.~Martinez and J.A.~Rodriguez, Phys. Rev. {\bf D65} 057301 (2002);
J.~Sjolin, J. Phys. {\bf G29} 543 (2003);
D.~Atwood {\it et~al.}, Phys. Rept. {\bf 347} 1 (2001);
Z.~Hioki and K.~Ohkuma, Eur.Phys.J. {\bf C65} 127 (2010).

\bibitem{chromo_top}
D.~Choudhury and P.~Saha, arXiv:0911.5016 [hep-ph].

\bibitem{CTEQ}
J.~Pumplin {\it et~al}, JHEP {\bf 0207} 012 (2002).

\bibitem{CDF_top_csec} 
CDF Public Note 9913 (http://www-cdf.fnal.gov/physics/new/top/2009/xsection/ttbar\_combined\_46invfb/).

\bibitem{CDF-D0} 
The Tevatron Electroweak Working Group, arXiv:0903.2503 [hep-ex].

\bibitem{Cacciari} 
M.~Cacciari {\it et~al.}, JHEP {\bf 0809} 127 (2008)

\bibitem{dim_6}
K.~Whisnant {\it et~al.}, Phys. Rev. {\bf D56} 467 (1997);
J.M.~Yang, B.L.~Young, Phys. Rev. {\bf D56} 5907 (1997).

\bibitem{CDF_mtt}
T.~Aaltonen {\it et~al.} (CDF Collaboration), Phys. Rev. Lett. {\bf 102}  222003 (2009).

\bibitem{CMS-PAS} 
CMS Physics Analysis Summaries (CMS-PAS-TOP-09-002, CMS-PAS-TOP-09-004, CMS-PAS-TOP-09-010), The CMS Collaboration\\ 
(http://cdsweb.cern.ch/collection/CMS\%20PHYSICS\%20ANALYSIS\%20SUMMARIES).

\bibitem{Rizzo_94}
T.G.~Rizzo, Phys. Rev. {\bf D50} 4478 (1994). 

\bibitem{Rizzo_96}
T.G.~Rizzo, arXiv:hep-ph/9605361 (1996). 

\end{thebibliography}
\end{document}